# The Mellon Fedora Project
## Digital Library Architecture Meets XML and Web Services

Sandra Payette[1] and Thornton Staples[2]

[1]Department of Computer Science, Cornell University
`payette@cs.cornell.edu`
[2] University of Virginia Library
`staples@virginia.edu`

**Abstract.** The University of Virginia received a grant of $1,000,000 from the Andrew W. Mellon Foundation to enable the Library, in collaboration with Cornell University, to build a digital object repository system based on the Flexible Extensible Digital Object and Repository Architecture (Fedora). The new system demonstrates how distributed digital library architecture can be deployed using web-based technologies, including XML and Web services. The new system is designed to be a foundation upon which interoperable web-based digital libraries can be built. Virginia and collaborating partners in the US and UK will evaluate the system using a diverse set of digital collections. The software will be made available to the public as an open-source release.

## 1 Introduction

In September of 2001 The University of Virginia received a grant of $1,000,000 from the Andrew W. Mellon Foundation to enable the Library, in collaboration with Cornell University, to build a sophisticated digital object repository system based on the Flexible Extensible Digital Object and Repository Architecture (Fedora) [1][2][3]. Fedora was originally developed as a research project at Cornell University, and successfully implemented at Virginia in 2000 as a prototype system to provide management and access to a diverse set of digital collections [4].

The Mellon grant was based on the success of the Virginia prototype, and the vision of a new open-source version of Fedora that exploits the latest web technologies. Virginia and Cornell have joined forces to build this robust implementation of the Fedora architecture with a full array of management utilities necessary to support it. A deployment group, representing seven institutions in the US and the UK, will evaluate the system by applying it to testbeds of their own collections. The experiences of the deployment group will be used to fine-tune the software in later phases of the project.

The motivation for the new system specification is to create an implementation of Fedora that is highly compatible with the web environment - one that uses web standards, and is built with freely available technologies. The original Fedora research implementation was built in a distributed object paradigm using the Common Object

Request Broker Architecture (CORBA). The Virginia reinterpretation proved that the model could be adapted to run as a web application, specifically using Java Servlet technology with relational database underpinnings. However, the prototype sacrificed some of the advanced interoperability features of Fedora. The new Mellon Fedora has been carefully designed to recreate a full-featured Fedora system that can become a foundation upon which interoperable web-based digital libraries can be built.

With the advent of XML and web services, a new paradigm for web-based applications is emerging. The new Mellon Fedora open-source system offers the opportunity to deploy interoperable digital libraries using mainstream web technology. The Fedora access and management systems are described using the Web Services Description Language (WSDL), as are all auxiliary services included in the architecture. The system communicates over HTTP and supports the Simple Object Access Protocol (SOAP). The project has adopted the Metadata Encoding and Transmission Standard (METS) [9] as the means to encode and store digital objects as XML entities.

This paper is a report on the status of the Mellon Fedora project. The design phase is complete and the detailed system specification is available from the project's SourceForge distribution site [10]. The alpha release of the software will be available for download by deployment partners in October 2002. We are planning a public release date of January 2003.

## 2   Virginia Requirements for Managing Large Digital Collections

The University of Virginia Library has been building digital collections since 1992. The Library has amassed a large collection that includes a variety of SGML encoded etexts, digital still images, video and audio files, and social science and geographic data sets that are being served to the public from a collection of independent web sites that have very little cross-integration. By 1999 it became clear that the Library's future involved very large-scale collections in all media and content types.

Like many other libraries, Virginia initially sought a vertical vendor solution that provided a complete, self-contained package for delivering and managing all digital content needs. A number of commercial solutions were considered, including IBM's Digital Library Software system (later renamed Content Manager) and SIRSI's Hyperion digital media archive system. The investigation started with the requirement for a digital content repository with a wide variety of features, including scalability to handle hundreds of millions of digital resources, flexibility to handle the ever expanding list of digital media formats, and extensibility to facilitate the building of customizable tools and services that can interoperate with the repository. It is clear that such repository functionality needs to be the core of a digital library system providing a means of uniquely identifying each piece of digital content as well as identifying groups of related content or collections. The remaining services and functionality of a digital library system would then be built on top of this core.

The Virginia search revealed a number of shortcomings in commercial digital library products:

- Most products are narrowly focused on specific media formats that offer good solutions for managing and delivering video or images but lack adequate tools and support for structured (e.g., XML or SGML) electronic texts or the ability to intermingle media types.
- Many products perform well at document management but offer no features for dealing with video or images.
- None of the products we examined adequately addressed the need to track and manage the array of ancillary programs and scripts that play an essential role in the delivery of that digital content.
- Many products fail to effectively deal with the complex interrelationships among digital content entities. As an example, consider an electronic text in the form of a five hundred-page book. The book consists of a single file containing all five hundred pages of text, marked up using XML. In addition to the XML file, there are also five hundred images that represent the scanned pages from the original hardcopy edition of the book. There are also twenty-five audio files that provide a recording of the book's content read aloud. To the librarian, all of these digital media are digital manifestations of the intellectual object known as the "book" and all are closely related to one another.
- Few of the products attended to the critical issue of interoperability, failing to provide an open interface to allow sharing services and content with systems from other vendors at other libraries.

In the summer of 1999, early in the design process, the Library's research and development group discovered a paper about Fedora written by Sandra Payette and Carl Lagoze of Cornell's Digital Library Research Group. Fedora was designed on the principle that interoperability and extensibility is best achieved by architecting a clean and modular separation of data, interfaces, and mechanisms (i.e., executable programs). With Cornell's help, the Virginia team installed the research software version of Fedora and began experimenting with some of Virginia's digital collections. Convinced that Fedora was exactly the framework they were seeking, the Virginia team reinterpreted the implementation and developed a prototype that used a relational database backend and a Java servlet that provided the repository access functionality. The prototype provided strong evidence that the Fedora architecture could indeed be the foundation for a practical, scalable digital library system.

## 3   Fedora: The basic architectural model

The Fedora architecture has been discussed at length elsewhere [1][2][3]. Similar architectures have also been described including Kahn/Wilensky [5], CNRI [6], Mönch [7], and Nelson [8]. We will review the basic Fedora architectural abstractions and introduce a slightly modified vocabulary to facilitate the discussion of the UVA prototype and the new Mellon implementation.

The two fundamental entities in Fedora are the *digital object* and the *repository*. As depicted in Figure 1, a digital object has a unique persistent identifier (*PID*), one or more `disseminators`, one or more `datastreams`, and system metadata. One significant characteristic of the Fedora digital object is that it aggregates both content (i.e., data and metadata) and behaviors (i.e., services); both can be distributed and referenced via a URI. As shown in Figure 1, `datastreams` represent content and `disseminators` represent services. A Fedora repository provides both access and management services for digital objects.

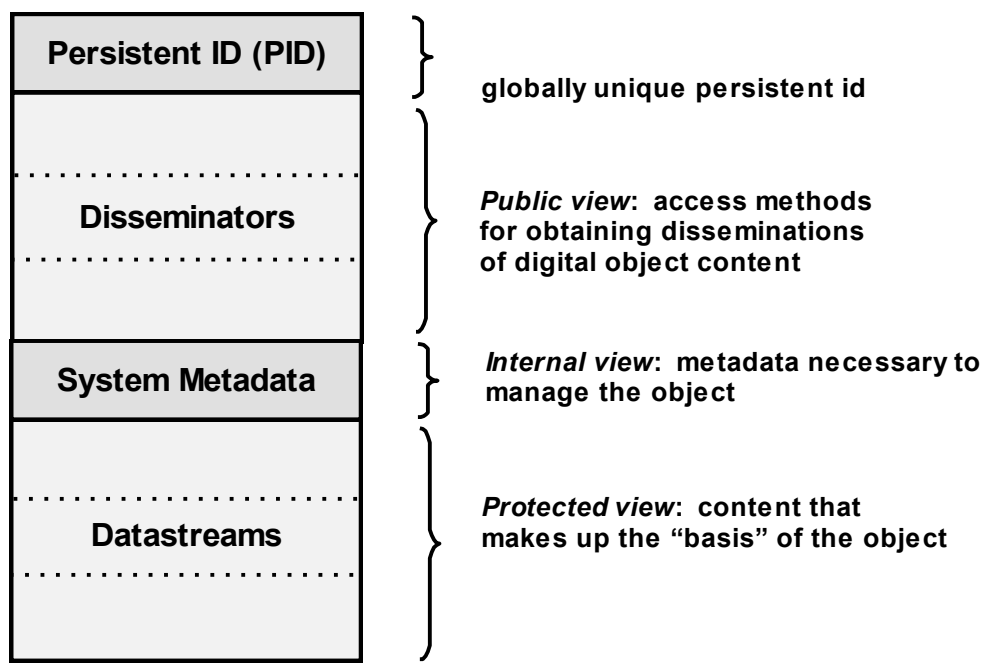

Figure 1 : Fedora Digital Object Model

From an access perspective, the architecture fulfills two basic functions: (1) it exposes both generic and extensible behaviors for digital objects (i.e., as sets of method definitions), and (2) it performs *disseminations* of content in response to a client's invoking these methods. A dissemination is defined as a stream of data that manifests a view of the digital object's content.

`Disseminators` are used to provide public access to digital objects in an interoperable and extensible manner. Object-specific access control policies can also be applied to `disseminators` [3]. Essentially, each `disseminator` will define a set of methods through which the object's `datastreams` can be accessed. For example, there are simple `disseminators` that define methods for obtaining different rendi-tions of images. There are more complex `disseminators` that define methods for interacting with complex digital creations such as multi-media course packages (e.g., GetSyllabus, GetLectureVideo). Finally, there are `disseminators` that define methods for transforming content (e.g., such as translating a text between different languages). A `disseminator` is said to "subscribe" to a `behavior definition`, which is an abstract service definition consisting of a set of methods for presenting or transforming the content of a digital object. A `disseminator` uses a `behavior mechanism`, which is an external service implementation of the methods to which the `disseminator` subscribes. A `disseminator` also defines the binding relationships between a `behavior mechanism` and `datastreams` in the object. The mechanics of how `disseminators` enable digital objects to interface with external services will be discussed later in Section 5.

## 4  The Virginia Testbed and Prototype

The initial goal of Virginia's prototype was to demonstrate a system that could re-create the same user experience that was currently being delivered through the Library's web site, but with the Fedora management and delivery architecture underneath. The next goal was to demonstrate how the Fedora architecture could enable alternative experiences of the original content.

As part of the prototype development process, Virginia focused on the design of different *content models*. A content model is a design pattern for a digital object - particularly the types of `datastreams` and `disseminators` in the object. One of the most important aspects of a digital object's design pattern is the definition of appropriate behaviors for the object. Figure 2 depicts two different content models developed for images at Virginia. Digital Object A contains four `datastreams`, one for each resolution of an image scan. Digital Object B contains one `datastream` for a wavelet-encoded image file. Despite these differences, both objects have `disseminators` that subscribe to the same `behavior definition`. This enables the image objects to be accessed from one abstract point of view (i.e., via methods like GetThumbnail and GetHighResolution). The objects are also said to have *functional equivalency,* which means that clients can interact with them in an interoperable manner.

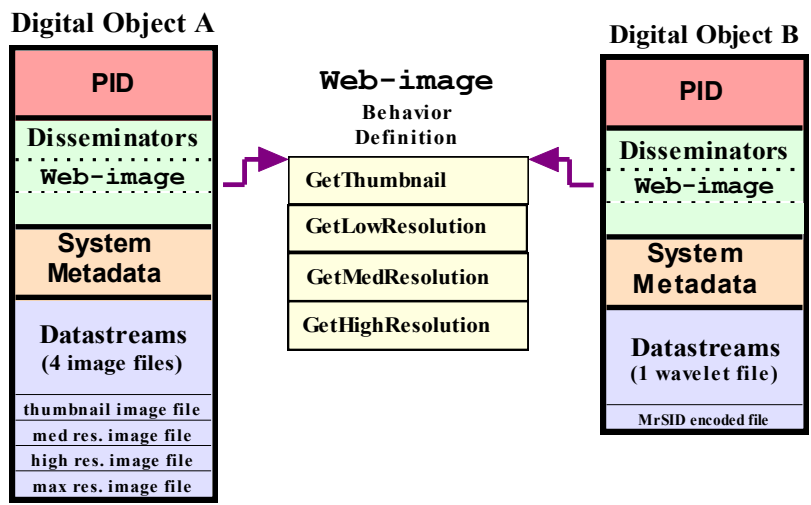

**Figure 2: Functional equivalency between digital objects**

Within a year, the Virginia testbed had grown to include 500,000 objects including digital images, numeric data, and XML objects. The team developed a variety of `disseminators` that provide functionality for electronic finding aids, TEI-encoded e-texts (letters and books), and structured XML-encoded art and archeology image collections. Another interesting `disseminator` provides access to a social science dataset. This object model includes `datastreams` to represent an XML codebook and an SQL dataset for a group of CBS news polls, with `disseminators` that allow investigation of the codebook and extraction/download of the data.

In addition to Library digital collections, the testbed included two "born-digital" projects created by humanities scholars in the Institute for Advanced Technology in the Humanities (IATH) at Virginia: the Salisbury Cathedral project and the Rossetti Archive. These two projects alone consist of approximately 6,000 digital images of art and architecture, and over 5,000 XML transcriptions of texts. The Supporting Digital Scholarship (SDS) project, jointly undertaken by the Library and IATH, focuses on collecting such projects in a Fedora repository.

The Virginia Fedora prototype was stress-tested using Apache's JMeter software [11]. Tests were conducted to simulate the load from a group of users simultaneously issuing a mixture of dissemination requests. These tests provided a simple proof of concept that the Fedora architecture can perform very efficiently in a production setting. On a Sun Ultra80 two-processor workstation, simulations of 20 simultaneous users, each making requests with an average delay of 300 milliseconds, yielded an average response time of approximately one half second per request. Note that most of the XML object transactions included a server-side rendering of XML into HTML, a relatively processor-intensive action. The repository was then moved to a four-processor, dedicated server, and the testbed was scaled up by repeatedly duplicating the existing 500,000 objects. Within a range of 1,000,000 to 10,000,000 objects, the same simulation yielded an average response time of approximately 1.5 seconds per transaction. We will use this data as the benchmark for performance in the new Mellon Fedora implementation.

## 5  The New Mellon Fedora: XML, Web Services, Versioning

Building on Virginia's extensive collections experience, the Virginia prototype and testbed, and the original Fedora research implementation, the Mellon Fedora team has developed an updated Fedora system specification that exemplifies a new generation of distributed web applications. Specifically, the original Fedora model has been reinterpreted using XML and Web services technologies. Our new implementation has the following key features:

- The Fedora repository system is exposed as a Web service and is described using Web Services Description Language (WSDL).
- Digital Object behaviors are implemented as linkages to distributed web services that are expressed using WSDL and implemented via HTTP GET/POST or SOAP bindings.
- Digital objects are encoded and stored as XML using the Metadata Encoding and Transmission Standard (METS).
- Digital objects support versioning to preserve access to former instantiations of both content and services.

Some brief definitions of Web service concepts are in order. In the most general sense, a Web service can be defined as a distributed application that runs over the internet. Web services are typically configured to use HTTP as a transport protocol for sending messages between different parts of the distributed application. The use of XML is a key feature of such applications, serving as a standard for encoding structured messages that are sent to and from the distributed applications. WSDL is an XML format for describing network services as a set of abstract operations that are realized as a set of *endpoints* that are able to receive and respond to structured messages [12]. Each endpoint communicates over a specific network protocol and uses a specific message format. There is currently some debate as to the preferred way to implement Web services. Some argue that messages can be exchanged in a simple manner over HTTP using GET and POST operations that return XML[13]. Others argue for the use of SOAP [14], which was originally conceived of as a way to do Remote Procedure Calls (RPC) with XML messaging. SOAP assumes no particular transport protocol, although it is common to send SOAP messages over HTTP.

The Mellon Fedora system will expose its services in both manners. An overview of the design of the new Fedora system is depicted in Figure 3. We will discuss the architecture diagram from the top down.

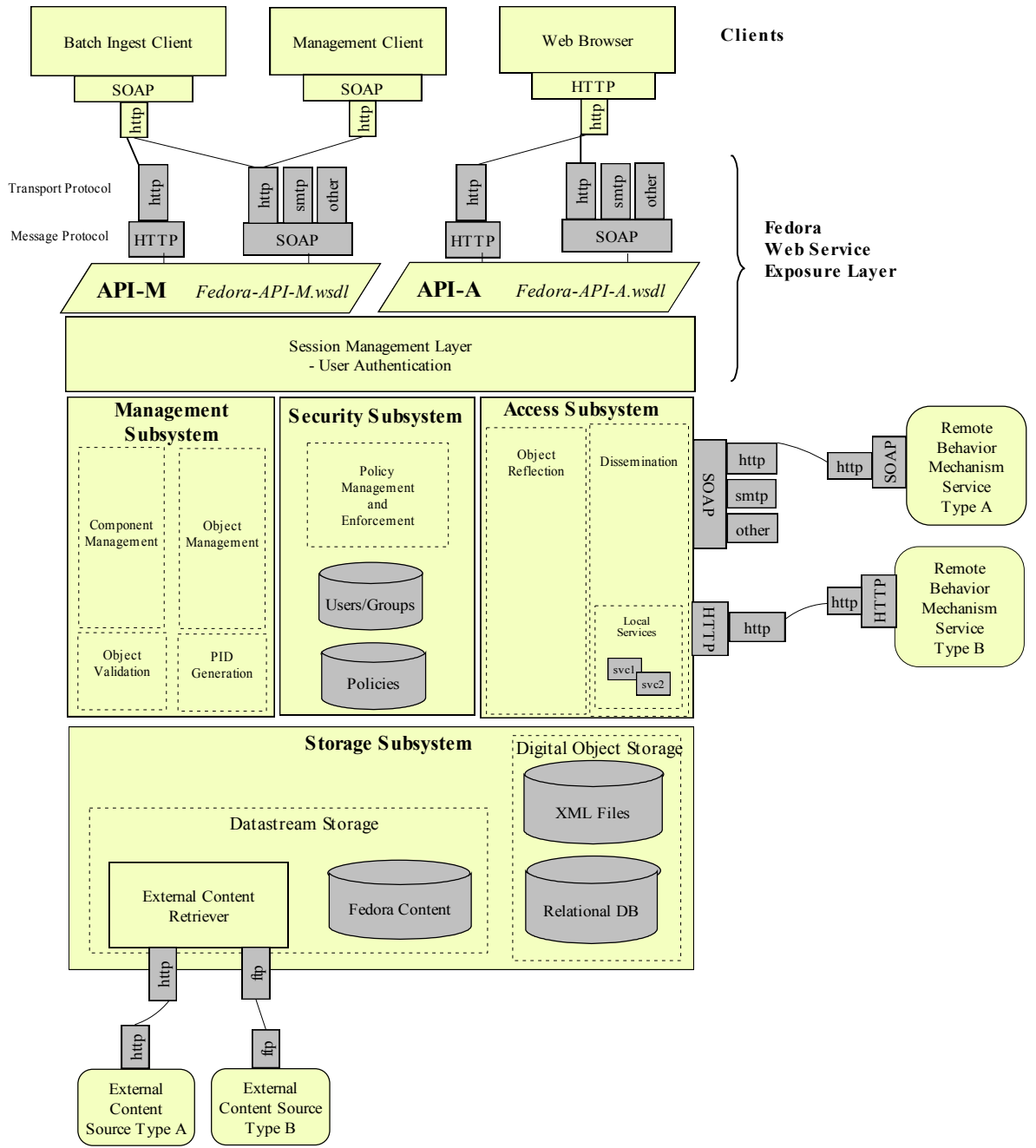

**Figure 3 : Mellon Fedora System Diagram**

### 5.1 Fedora Web Service Exposure Layer

The new Fedora system is exposed as two related web services, the Fedora Management service (API-M) and the Fedora Access service (API-A). The service interfaces are expressed in XML using WSDL. The Fedora Management service defines an open interface for administering the repository, including creating, modifying, and deleting digital objects, or components within digital objects. The Fedora Access service defines an open interface for accessing digital objects. The access operations include methods to do reflection on a digital object (i.e., to discover the kinds of disseminations that are available on the object), and to request disseminations. Clients can interact with the repository services either via HTTP or SOAP. The WSDL for each repository service defines bindings for both modes of communication. In Phase I of the project, the HTTP GET/POST service bindings will connect to a Java Servlet running on Apache Tomcat, and SOAP bindings will connect to a SOAP-enabled web service running on Apache Axis with Tomcat.

### 5.2 Access Sub-System

The Access sub-system supports digital object reflection and disseminations of digital object content. It implements three basic operations described in the Access service definition:

- *GetBehaviorDefTypes* - identifies the types of `behavior definitions` the object subscribes to.
- *GetMethods* - returns a WSDL description of the methods for a particular `behavior definition`
- *GetDissemination* - runs a method on the digital object to produce a dissemination

To better understand the Access sub-system, we will first examine the digital object model from the perspective of how `disseminators` are used to associate access methods with an object. To review, a digital object aggregates content in the form of `datastreams`, and assigns behaviors (access methods) in the form of `disseminators`. A `disseminator` references an abstract definition of a set of methods and a mechanism (service) for running those methods. When clients issue dissemination requests for a behavior method, supporting services are called to release `datastreams` from the object, or provide transformations of the `datastreams`.

*Linkages to supporting services*

How does all of this work if neither the method definitions nor the service implementations actually reside within a digital object? Basically, the Fedora Access sub-system acts as a service mediator for clients accessing digital objects.

In Figure 4, a digital object has an Image Watermarker `disseminator`. The `disseminator` has two notable attributes: a `behavior definition` identifier and a `behavior mechanism` identifier. These identifiers are actually *persistent identifiers to other Fedora digital objects*. These are special digital objects that are *surrogates* for external services, for example, a service for obtaining watermarked images at different resolutions. A `behavior definition` object contains a special `datastream` whose content is a WSDL definition of abstract methods for images (e.g., GetThumbnail, GetWatermarked). A `behavior mechanism` object contains a special `datastream` that is a WSDL definition describing the run-time bindings to an external service for these methods (operations). Service bindings can be via HTTP GET/POST or SOAP.

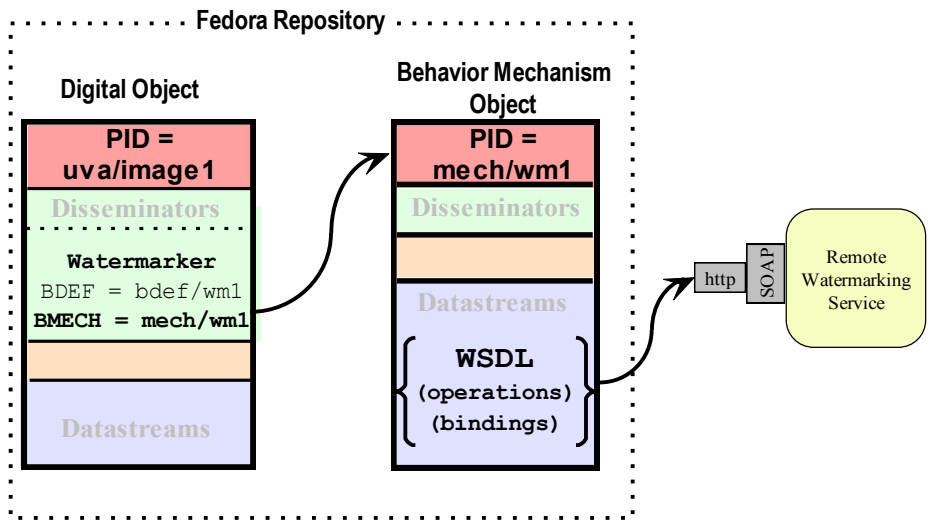

**Figure 4 : Digital Object Association to External Service**

Thus, a key function of the Fedora Access sub-system is to fulfill a client's request for dissemination by evaluating the behavior associations specified in a digital object, and figuring out how to dispatch a service request to an external service with which the digital object associates. The Access sub-system facilitates all external service bindings on behalf of the client, simply returning a dissemination result. Note that a client can be a web browser, a web application with embedded dissemination requests, or a custom client built to interact with Fedora.

### 5.3 Management and Security Sub-Systems

The Management sub-system implements an array of operations for creating and maintaining digital objects. This sub-system mediates the creation and manipulation of XML-encoded digital objects, in response to service request from clients via API-M. The component management module is responsible for maintaining version control within digital objects, as described in Section 7. The validation module ensures that each operation performed on a digital object is valid from the METS schema perspective. It also validates objects against a set of Fedora-specific integrity rules, especially referential integrity between digital objects and the behavior service entities to which they refer. A PID generation module is responsible for dispensing unique persistent identifiers for digital objects.

It should be noted that the object management module provides an ingestion function that will accept METS-encoded digital objects created outside the context of the repository system. This facilitates batch loading of digital objects and movement of objects among repositories. All ingested objects are subject to the integrity constraints enforced via the validation module. A full description of the management operations and their implementation details can be obtained from the Mellon Fedora project website [10].

The Security sub-system enables repository managers to define access control policies for the repository. It also provides the mechanism to enforce these policies at runtime. In Phase I, the basic repository management functions (API-M) will be secured through a technique known as Inline Reference Monitoring [15] using the Policy Enforcement Tookit (PoET) developed at Cornell[16]. PoET provides a highly expressive policy enforcement language that will be used to define a repository-wide policy to prevent unauthorized users from performing secured tasks. PoET's enforcement scheme involves dynamically modifying java bytecode modules, infusing the application with the policy rules at runtime. This policy enforcement scheme was successfully demonstrated in the research implementation of Fedora at Cornell [3]. In Phase, II we will focus on XML-oriented policy expression and the enforcement of fine-grained object-level policies.

### 5.4 Storage Sub-System

The Storage sub-system manages all aspects of reading, writing, and deleting data from the repository. Within it we find the actual data store for the XML-encoded digital objects, and the `datastreams` to which the digital objects refer.

Digital objects are stored as XML files. All digital objects conform to the METS XML schema, as described in Section 6. Digital object XML files aggregate one or more `datastreams`. `Datastreams` are of two types: those under the custodianship of the repository system (i.e., Fedora content ) and those that are references to external content (i.e., represented as a URI to a remote content source). The Storage sub system is responsible for managing content files stored within its domain, and for interfacing with remote content sources to obtain external content at runtime. The Phase 1 implementation supports retrieval of external content via back-end HTTP and FTP; however, in later phases, other protocol gateways will be introduced. For example, we envision `datastreams` that are stored SQL queries, stored SOAP requests, and even stored dissemination request to other Fedora digital objects.

The Phase I Storage sub-system supports a relational database as an alternate form of storage for digital objects. This is a redundant storage scheme that enables us to guarantee the performance benchmark set by the Virginia prototype. The team created a relational database schema that expresses a Fedora digital object from an access point of view. The Storage sub-system is responsible for replicating the XML-encoded digital objects into the database schema format.

The relational database is a *temporary* feature that will service disseminations while we optimize performance for querying the XML-based digital objects.

## 6  Encoding Digital Objects using METS

A major goal of the new Mellon Fedora project was to define an XML schema for the Fedora digital object model. Early in the project, the team discovered the Metadata Transmission and Encoding Standard (METS), a Digital Library Federation [17] initiative focused on developing an XML format for encoding metadata necessary to manage digital library objects within a repository and to facilitate exchange of such objects among repositories. METS is expressed using the XML Schema language [18] and is freely available from the METS website [9]. The METS standard is maintained by the Network Development and MARC Standards Office of the Library of Congress.

From the Fedora perspective, the METS schema provided much of the functionality required to encode digital objects. However, the concept of associating behaviors or services with objects was not initially supported by METS. The Fedora project joined the METS specification effort, and was instrumental in effecting additions to the METS schema to support the Fedora notion of a `disseminator`. The Mellon Fedora project now uses METS as the official encoding format for digital objects stored in a Fedora repository.

All major components of a Fedora digital object can be mapped to elements defined in the METS schema. The relationships among digital object components are also easily expressed in METS. Table 1 shows the translation of the major Fedora digital object components to their equivalent METS entities. The XML samples in the METS column are abbreviated for readability.

Table 1: Mapping Fedora Digital Object to METS

| Fedora | METS Encoding |
| --- | --- |
| **Persistent Identifier** | <METS:mets OBJID=**"*PID*"** /> |

| System Metadata | `<METS:metsHdr/>`<br>`<METS:amdSec/>` |
|---|---|
| **Datastreams**<br><br>*Fedora User Metadata*<br><br>*Fedora Content*<br>+<br>*External Referenced Content* | `<METS:amdSec/>`<br>`<METS:dmdSec/>`<br><br>`<METS:fileGrp ID="`**dsID**`"`<br>`   <METS:file ID="`**dsVersionID**`>`<br>`      <METS:Flocat LOCTYPE="URL" xlink:href="`**dsLocation**`"/>`<br>`   </METS:file>`<br>`</METS:fileGrp>` |
| **Disseminators**<br><br>*Behavior Definition*<br>+<br>*Service*<br><br>*Datastream Relationships* | `<METS:behaviorSec ID="`**dissID**`" STRUCTID="`**dsMapID**`">`<br>`   <METS:interfaceDef  xlink:simpleLink="`**bdefID**`"/>`<br>`   <METS:mechanism xlink:simpleLink="`**bmechID**`" />`<br>`</METS:behaviorSec>`<br><br><br>`<METS:structMap ID="`**dsMapID**`">`<br>`   <METS:div TYPE="`**dsBindName**`" ORDER="`**dsSeq**`">`<br>`      <METS:fptr FILEID="`**datastreamID**`" />`<br>`   </METS:div>`<br>`</METS:structMap>` |

## 7  Digital Object Versioning Strategy

The Mellon Fedora system supports versioning within digital objects to preserve former instantiations of content and services. Specifically, the system creates versions of `datastreams` and `disseminators` within a digital object. From a management perspective this provides a mechanism to track changes in objects over time. From an Access perspective, this enables users to view digital objects from a historical perspective. As previously mentioned, the Management sub-system will be responsible for the versioning task. Rather than maintaining multiple instances of digital object XML files, the system will maintain versions of `datastreams` and `disseminators` *within* a digital object. The system will also insert audit trail records describing the changes. This strategy enables individual object components to evolve at their own pace, and provides a container for the entire object history. The XML snippet below is an example of how `datastream` versioning is encoded using

METS. The METS <fileGrp> is used to group versions of the same `datastream`. Each `datastream` version will be represented by a METS <file> element. The ID attributes of the <file> element will be used for version numbers. The CREATED attribute of the <file> provides the date and time that a version was created.

```xml
<METS:fileGrp ID=Datastreams>
    <METS:fileGrp ID="DS1">
        <!--This is the most current version of the high-resolution image -->
        <METS:file ID="DS1.1" SEQ="1" CREATED="2002-08-31T06:32:00"
                MIMETYPE="image/jgp" ADMID="audit2">
            <METS:FLocat LOCTYPE="URL" xlink:href="http://uva.edu/img1.jpg"/>
        </METS:file>
        <!--This is an OLDER version of the high-resolution image -->
        <METS:file ID="DS1.0" SEQ="2" CREATED="2002-01-22T06:32:00"
                MIMETYPE="image/jgp" ADMID="audit1">
            <METS:FLocat LOCTYPE="URL" xlink:href="http:// uva.edu/img1a.jpg"/>
        </METS:file>
    </METS:fileGrp>
</METS:fileGrp>
```

Versioning to reflect changes in behavior services is a more difficult task, since it can involve versioning of `disseminators`, as well as components within `behavior definition` and `mechanism` objects. We have designed a strategy to record changes such as the addition of new methods to a service definition, as well as major upgrades to a service implementation (i.e., a better mousetrap). The details of this scheme are described in the Mellon Fedora specification document available at the project website [10]. Again, the net effect of the digital object versioning strategy is to enable clients to obtain disseminations that reflect how a digital object looked at different points in time. To facilitate this, the Fedora Access service will support a date-time-stamped variant of the GetDissemination request.

## 8 Software Release Plan and Deployment Partnerships

The Phase 1 goals of the Mellon Fedora project are to publish the full system specification and to deploy the first open-source release the new Fedora software. The specification document has been published and can be accessed on the project web site. In October 2002, the team will release the alpha version of the software package for use by the deployment partners. A public release of software will be available on the project website in January 2003.

Phase 2, will entail working closely with our partners in testing and evaluating the system. Their experience will inform the development of subsequent releases of the software. The partners include: the Digital Library group at Indiana University; the Humanities Computing group at New York University; the Digital Collections and Archives Department at Tufts University; the Humanities Computing group at Kings

College, London; the Oxford Digital Library group and the Refugee Studies Center, both at Oxford University; and the Motion Picture Broadcasting and Recorded Sound Division at the Library of Congress; and a library/academic computing team from Northwestern University. Although not officially part of the deployment partnership, two other Mellon-funded projects underway at the University of Virginia will provide additional test implementations of the new Fedora system. The Digital Imprint project at the University of Virginia Press plans to experiment with Fedora as a means of publishing of born-digital scholarly projects. The Library's American Studies Information Community project, one of the seven Open Archives Initiative (OAI) projects recently funded by Mellon, will integrate American Studies information harvested from OAI servers into the Library's Fedora repository.

As previously mentioned, Phase 2 will entail performance optimization, especially for XML querying. This will be an important prerequisite for phasing out the supplemental relational database storage scheme for digital objects. Security and access control will be significantly enhanced. Interoperability experiments among deployment partners will be conducted to realize the full goals of the distributed Fedora system. The final phase will concentrate on providing more sophisticated delivery of end-user experiences. This will include extending the functionality of `disseminators` by adding new services that are important for collecting scholarly projects and publications. The project team also plans to map the major functions of the Fedora system to the OAIS reference model [19] and to develop a strategy for managing the open source software over time.

## 9  Concluding Statement

The Fedora collaboration between University of Virginia and Cornell University has been a model for moving digital library research into a production environment that is motivated by the needs of scholars, librarians, and other information communities. The Mellon project serves as an example of how to bridge computer scientists doing digital library research and institutions that are building large digital collections. The cycle of research, reference implementation, technical transfer, prototyping, redesign, and production implementation has ensured that the Mellon Fedora project is grounded in the requirements of real collections and users.

## 10  Acknowledgements

The authors would like to give special thanks to Ross Wayland for his keen analysis and leadership. Also, we wish to acknowledge the Fedora development team members for their excellent contributions to the design and implementation of the new system. The team includes (alphabetically): Paul Charlton, Ronda Grizzle, Carl Lagoze, Bill Niebel, Tim Sigmon, Ross Wayland, Chris Wilper. We would also like to thank Jerry McDonough of NYU for his support regarding METS. Finally, we are very grateful to the Andrew W. Mellon Foundation for its generous grant which has made this project possible.